\documentclass[%
 reprint,
 amsmath,amssymb,
 aps,
]{revtex4-2}
\usepackage[colorlinks,linkcolor=blue,anchorcolor=blue,citecolor=blue]{hyperref}
\usepackage{graphicx}
\usepackage{dcolumn}
\usepackage{bm}
\usepackage{float}
\usepackage{multirow}
\begin{document}

\preprint{APS}

\title{Bell state analysis using orbital angular momentum and path degrees of freedom}
\author{Zi-Long Yang,$^{1}$ Shi-Wen He,$^{1}$ Lin-Cheng Wang,$^{1}$ Si-Tong Jin,$^{1}$ Liu Lv,$^{3}$ Xiao-Ming Xiu,$^{2}$ Chong Li$^{1}$}
\email{Corresponding author: lichong@dlut.edu.cn}
\affiliation{$^{1}$School of Physics, Dalian University of Technology, Dalian 116024, China\\$^{2}$College of Physical Science and Technology, Bohai University, Jinzhou 121013, China\\$^{3}$College of Mechanical Engineering, Chaoyang Normal University, Chaoyang 122000, China}
\date{\today}

\begin{abstract}

Bell state analysis (BSA) constitutes a foundational operation for distinguishing Bell states in numerous quantum information processing (QIP) protocols. In this work, we propose a theoretical scheme for realizing a perfect BSA tailored for polarized Bell states, with assistance from orbital angular momentum (OAM) and path entanglement. The linear-optics-based architecture for BSA circumvents the inherent limitations of nonlinear optical processes and enhances the robustness against environmental noise\textemdash a major challenge in practical QIP implementations. The integrating hyperentanglement (combining polarization, OAM, and path degrees of freedom (DOFs)) raises the theoretical success probability to 100\%, achieving deterministic BSA. This deterministic BSA scheme offers a promising route toward practical, high-performance QIP in photonic systems, leveraging current experimental techniques and addressing key limitations of existing methods.

\end{abstract}

\maketitle


\section{\label{sec:level1} INTRODUCTION }

Bell states~\cite{Einstein1935pr777, Bell1964prf195, Zhang2016nc13514}, as fundamental resources for QIP tasks (e.g., quantum communication~\cite{Qi2025pra052609, An2025prl010804}, quantum computation~\cite{Hang2024prl020601, Hauser2025npj41}, and quantum measurement~\cite{Jia2025sci3576, Virz2024qst045027}), have been experimentally realized across diverse quantum platforms~\cite{Sarika2024po100199, LIU2022Chip100001, Hong2025pra034038, Shankar2013n419, Leghtas2013pra023849, Xu2024qst035002, Guimond2019prl093601, Clark2021prl130505, Hu2022pra062422, Neg2018n527, Fehler2020np3655, Aman2020pe113915, Hu2024nc10256, Cai2024np1022}. In photonic systems, Bell states are generated by applying linear operations to entangled photon pairs produced via spontaneous parametric down conversion~\cite{Sarika2024po100199, LIU2022Chip100001, Hong2025pra034038}. For superconducting quantum systems, Bell states are produced by engineering entanglement between independent superconducting qubits using microwave pulses~\cite{Shankar2013n419, Leghtas2013pra023849, Xu2024qst035002}. In atomic systems, precise control of atomic interactions with laser pulses enables the realization of Bell states~\cite{Guimond2019prl093601, Clark2021prl130505, Hu2022pra062422}. Beyond these mainstream platforms, analogous implementations have been demonstrated in a variety of other quantum platforms~\cite{Neg2018n527, Fehler2020np3655, Aman2020pe113915, Hu2024nc10256, Cai2024np1022}.

While the experimental realization of Bell states across these diverse quantum platforms lays the foundation for practical QIP implementations, the full exploitation of their entangled nature hinges on a complementary and equally critical capability\textemdash precise discrimination of the four distinct Bell states, a task formally referred to as BSA~\cite{Weinfurter1994EL559, Michler1996praR1209, Kwiat1998praR2623}. Complete BSA enables the accurate extraction and utilization of quantum information encoded in entangled states. is indispensable for a wide range of core QIP protocols, including quantum teleportation~\cite{Bouwmeester1997n575, Bianchi2025pra012615, kin2025prl160802}, quantum key distribution~\cite{Kovalenko2021pr2351, Le2025PRX020311, Tagliavacche2025npj04}, quantum dense coding~\cite{Zhang2005SCPMA57, Piveteau2022nc2041, Berthusen2025PRX010306}, and quantum entanglement swapping~\cite{Pan1998prl3891, Jing2022prl060503, Li2025aqt2500157}. However, a well-established fundamental constraint in linear optics is that complete BSA cannot be achieved with linear optical elements alone, and the maximum success probability is capped at 50\%~\cite{Walborn2003el161, REN2005pla8, Walther2005pra010302, Wei2007pra060305, Schuck2006prl190501, Grice2011pra042331, Fields2022pra023706, Matthias2023sa4080, Li2024lsa07}. To circumvent this constraint, various schemes have been proposed using auxiliary photons~\cite{Calsamiglia2001apb67, DUSEK2001oc161, Zhou2016sr20901}, nonlinear optical systems~\cite{Ren2012oe24664, Sheng2015sr13453, Liu2016sr22016, Li2016OE18388, Kozubov2021SR14236}, or auxiliary atoms~\cite{Hans2005sci586, Ren2013tepjd30, Wang2016OE28444, Wang2016sr19497, Zeng2018lpl055204, Cao2020OE2857}. Unfortunately, these approaches suffer from inherent drawbacks. Auxiliary photons tend to introduce environmental noise and increase quantum circuit complexity~\cite{Calsamiglia2001apb67, DUSEK2001oc161, Zhou2016sr20901}. Nonlinear optical systems complicate photon manipulation and compromise optical path stability~\cite{xiu2016pra042321, Li2016OE18388, Kozubov2021SR14236}. Auxiliary atoms exhibit limited coherence times, which hinders the development of compact and scalable QIP units~\cite{Yang2024aqt2300376, Dong2025olt112583, Chen2025aqt00534}.

To mitigate the limitations of BSA schemes relying on extra quantum resources (e.g., auxiliary photons or atoms), researchers have proposed linear optical approaches by harnessing new DOFs of quantum systems\textemdash an elegant strategy to avoid introducing external resources while enhancing BSA performance~\cite{Li2017pra020303, Wang2019oe8994, Zhang2019pra052301, Zeng2020lpl075203, Gao2020ape027004, Wang2022oe34468, Zeng2022ol5817, Zhou2022fp41502, Fan2024qip316, Zeng2025josab1949}. Li et al. and Wang et al. independently proposed linear optical hyperentangled BSA (HBSA) schemes encoded in polarization and longitudinal momentum entanglement, with an additional longitudinal momentum DOF serving as the auxiliary~\cite{Li2017pra020303, Wang2019oe8994}. Zhang et al. designed a high-dimensional linear optical BSA scheme entangled in the path DOF using the auxiliary OAM DOF~\cite{Zhang2019pra052301}. Assisted by time-bin–frequency hyperentanglement, Zeng et al. present another HBSA scheme encoded in polarization and spatial-mode DOFs in linear optical systems~\cite{Zeng2020lpl075203}. Gao et al. proposed a universal linear optical protocol capable of fully distinguishing 16 hyperentangled Bell states entangled in polarization and momentum DOFs with the aid of the time-bin DOF~\cite{Gao2020ape027004}. Zhou et al. proposed a linear optical HBSA scheme encoded in spatial and polarization DOFs assisted by a fixed frequency-based entanglement and a time interval DOF~\cite{Zhou2022fp41502}. For four-dimensional OAM-Bell states, complete high-dimensional BSA schemes have been achieved in linear optical systems via distinct routes, including auxiliary path DOF~\cite{Zeng2022ol5817}, auxiliary polarization DOF~\cite{Wang2022oe34468}, auxiliary frequency DOFs~\cite{Fan2024qip316}, or two-photon interference~\cite{Zeng2025josab1949}. Collectively, integrating auxiliary DOFs into linear optical systems not only enhances the discriminative capability for complete BSA, but also ensures high success probability while minimizing system complexity. By avoiding external auxiliary resources, these linear-optics-based BSA protocols lay a critical foundation for the scalable and compact implementation of quantum information technologies.

In this paper, we propose a theoretical two-dimensional BSA scheme in a linear optical system, which exploits the OAM and path as auxiliary DOFs. By performing single-photon projective measurements (SPPM) on auxiliary DOFs, the quantum information encoded in the Bell states can be deterministically identified without relying on nonlinear optical processes or additional quantum resources. This approach not only enhances the discrimination efficiency and scalability of BSA but also establishes a feasible route toward fully deterministic entanglement manipulation in photonic quantum networks. Furthermore, the proposed framework exhibits inherent compatibility with existing linear optical architectures, offering a pragmatic and experimentally accessible pathway for realization.

\section{Manipulation of states}
\subsection{Polarization-controlled OAM shift gate}
The schematic diagram of a polarization-controlled OAM shift (P-COS) gate is shown in Fig.~\ref{fig:2}(a). The P-COS gate consists of two quarter-wave plates oriented at an angle $-\pi/4$ [${\rm{QWP}(-\pi/4})$] and a q-plate with a topological charge $q$ [QP(q)]. The ${\rm{QWP}(-\pi/4})$ converts the linear (circular) polarization state to the circular (linear) polarization state~\cite{Xavier2020ph53, Nouman2016sr39062, Shah2019ejp065301}, and its operation on the polarization mode can be expressed as

\begin{eqnarray} 
\left| H,l \right\rangle \xrightarrow{{\rm{QWP}(-\pi/4)}} \left| L,l \right\rangle,
\left| L,l \right\rangle \xrightarrow{{\rm{QWP}(-\pi/4)}} \left| V,l \right\rangle,\nonumber\\
\left| V,l \right\rangle\xrightarrow{{\rm{QWP}(-\pi/4)}} \left| R,l \right\rangle,
\left| R,l \right\rangle\xrightarrow{{\rm{QWP}(-\pi/4)}} \left| H,l \right\rangle,
\end{eqnarray}
where $\left| H \right\rangle$ and $\left| V \right\rangle$ represent the horizontal and vertical polarization states, $\left| R \right\rangle$ and $\left| L \right\rangle$ represent the left and right circular polarization states, and $l$ is the OAM quantum number that taking integer value. The QP enables a perfect conversion of spin angular momentum (SAM) to OAM without transferring the angular momentum to the plate itself~\cite{Marrucci2006prl163905, Vert2017oe5719, PILKA2024olt110019, Pereiro2023sr202302}. When a circular polarized photon traverses a QP with topological charge $q$, its SAM is reversed, and its OAM is changed by $\pm 2q$. The action of ${\rm{QP}}(q)$ in the polarization and OAM modes can be described by
\begin{eqnarray}\label{qp}
\left| L,l \right\rangle \xrightarrow{{\rm{QP}}(q)} \left| R,l+2q \right\rangle,
\left| R,l \right\rangle \xrightarrow{{\rm{QP}}(q)} \left| L,l-2q \right\rangle.
\end{eqnarray}

In the implementation of the P-COS gate controlled by horizontal and vertical polarization modes, the polarization modes act as the control qubit, while the OAM modes act as the target qubit. That is, the polarization state is in horizontal mode $\left| H \right\rangle$ (vertical mode $\left| V \right\rangle$), its OAM mode is changed by $2q$ ($-2q$). Thus, the operation of the P-COS gate can be expressed as
\begin{equation}\label{qpq}
\left| {H,l} \right\rangle  \xrightarrow{\rm{P-COS}} \left| {H,l+2q} \right\rangle ,\ \left| {V,l} \right\rangle  \xrightarrow{\rm{P-COS}} \left| {V, l-2q} \right\rangle.
\end{equation}

\subsection{OAM-controlled path shift gate}
\begin{figure*}[htbp]
\centering
\includegraphics[scale=0.7]{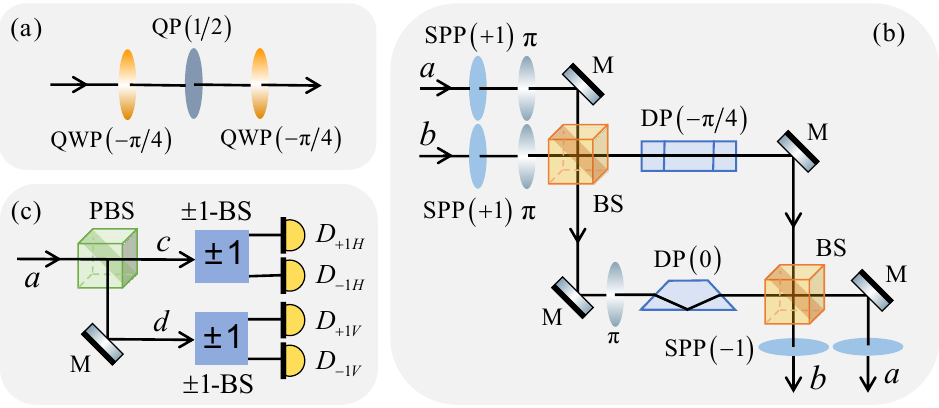}
\caption{\label{fig:1} Schematic diagram of quantum logic gates implemented using linear optical elements. (a) The P-COS gate is used to introduce a topological charge $2q$ controlled by horizontal and vertical polarization states. (b) The O-CPS gate is used to sort incident photons into distinct path modes determined by the positive and negative OAM modes. M denotes a mirror whose operation $\left| l \right\rangle  \to e^{i\pi/2}\left| -l \right\rangle $. (c) The SPPM is used to identify the output hyperentangled Bell states based on the clicks of different detectors.}
\end{figure*}

The schematic diagram of an OAM-controlled path shift (O-CPS) gate is shown in Fig.~\ref{fig:1}(b). The O-CPS gate consists of spiral phase plates (SPPs)~\cite{Wang2017Optica1467} and a Mach-Zehnder interferometer (MZI)~\cite{Yu2016cpb020304, Jia2017apl111, Salman2021np125}. The SPP$({l})$ changes an OAM quantum number of $l$. The MZI is composed of ${\pi}$ phase plate [PP$({\pi})$], 50:50 beam splitters (BSs) and dove prisms (DPs)~\cite{Gao2019pra023825, SUN2024olt110819}. The PP$({\pi})$ and DP implement the transformation $\left| l \right\rangle  \to i\exp \left( {i2\alpha l} \right)\left| { - l} \right\rangle$ and $\left| l \right\rangle  \to e^{i\pi}\left| l \right\rangle $ in OAM mode, respectively. The 50:50 BS allows an incident photon to be transmitted and reflected with equal probabilities of 50\%. 





For clarity, we consider an input state
\begin{eqnarray}
{\left| {\phi _{in} } \right\rangle}=\frac{1}{{\sqrt 2 }}({\left| p,+1 \right\rangle+\left| p,-1 \right\rangle})\left| a \right\rangle
\end{eqnarray}
injected into the O-CPS gate through the path $a$, where $p\in\{H, V\} $. The SPP$({+1})$ operations are performed in OAM modes, the input state ${\left| {\phi _{in} } \right\rangle}$ is changed to 
\begin{eqnarray}
{\left| {\phi _1 } \right\rangle}=\frac{1}{{\sqrt 2 }}({\left| p,+2 \right\rangle+\left| p,0 \right\rangle})\left| a \right\rangle.
\end{eqnarray}
The state ${\left| {\phi _1 } \right\rangle}$ is split by a BS into the two arms of the MZI after passing through the PPs$({\pi})$. Then two DPs are placed in each arm to introduce a relative phase shift $\alpha=\pi /2$, with the optical axis of one DP rotated by $\alpha=\pi/4$ relative to the other DP. After recombining at the second BS, the state ${\left| {\phi _1 } \right\rangle}$ is transformed into
\begin{eqnarray}
{\left| {\phi _2 } \right\rangle}=\frac{1}{{\sqrt 2 }}({\left| p,+2 \right\rangle\left| a \right\rangle+\left| p,0 \right\rangle}\left| b \right\rangle).
\end{eqnarray}
The SPP$({-1})$ operations are performed in OAM modes, which transform the state ${\left| {\phi _2 } \right\rangle}$ into the output state
\begin{eqnarray}
{\left| {\phi _{out} } \right\rangle}=\frac{1}{{\sqrt 2 }}({\left| p,+1 \right\rangle\left| a \right\rangle+\left| p,-1 \right\rangle}\left| b \right\rangle).
\end{eqnarray}
Analogously, for the input quantum state 
\begin{eqnarray}
{\left| {\phi' _{in} } \right\rangle}=\frac{1}{{\sqrt 2 }}({\left| p,+1 \right\rangle+\left| p,-1 \right\rangle})\left| b \right\rangle
\end{eqnarray}
injected into the O-CPS gate through the path $b$, the corresponding output state is given by
\begin{eqnarray}
{\left| {\phi' _{out} } \right\rangle}=\frac{1}{{\sqrt 2 }}({\left| p,+1 \right\rangle\left| b \right\rangle+\left| p,-1 \right\rangle}\left| a \right\rangle).
\end{eqnarray}

In the O-CPS gate, the path flip is controlled by positive and negative OAM modes, with the OAM modes serving as the control qubit and the path modes serving as the target qubit. The photon with positive OAM remains in the same path, while a photon with negative OAM is directed to the opposite path. The operations of the O-CPS gate can be described as
\begin{eqnarray}\label{ocnot}
&\left| +1 \right\rangle \left| a \right\rangle  \xrightarrow{\rm{O-CPS}} \left| +1 \right\rangle \left| a \right\rangle ,\nonumber\\
&\left| { - 1} \right\rangle \left| a \right\rangle  \xrightarrow{\rm{O-CPS}} \left| { - 1} \right\rangle \left| b \right\rangle,\nonumber\\
&\left| +1 \right\rangle \left| b \right\rangle  \xrightarrow{\rm{O-CPS}} \left| +1 \right\rangle \left| b \right\rangle ,\nonumber\\
&\left| { - 1} \right\rangle \left| b \right\rangle  \xrightarrow{\rm{O-CPS}} \left| { - 1} \right\rangle \left| a \right\rangle.
\end{eqnarray}


\subsection{OAM Hadamard gate}

\begin{table*}[htbp]
\centering
\caption{\label{tab:1}Detailed parameter configurations of the linear optical elements. Implementation of the OH gate using a sequence of a QP, a Hologarm, a QWP, HWPs, and a PBS, arranged from left to right. The initial states are in the first column on the left-hand side, while the corresponding output states appear in the last column on the right-hand side. The symbol ``--" represents no element.}
\begin{ruledtabular}
\begin{tabular}{cccccccc}
 Input state&QP&SPP$(l)$&QWP&HWP&HWP&PBS&Output state\\ \hline
 $\left|{H,+1}\right\rangle$&$1/2$&$-1$&$\rm{-\pi}/4$&$\rm{\pi}/8$&--&$\left| H \right\rangle \left\langle H \right|$&$\frac{1}{{\sqrt 2 }}\left( {\left| {H,+1} \right\rangle  + \left| {H, - 1} \right\rangle } \right)$\\
 $\left|{V,+1}\right\rangle$&$1/2$&$-1$&$\rm{-\pi}/4$&$\rm{\pi}/8$&--&$\left| V \right\rangle \left\langle V \right|$&$\frac{1}{{\sqrt 2 }}\left( {\left| {V,+1} \right\rangle  + \left| {V, - 1} \right\rangle } \right)$\\
 $\left|{H,-1}\right\rangle$&$1/2$&$+1$&$\rm{-\pi}/4$&$\rm{\pi}/8$&$\rm{\pi}/4$&$\left| H \right\rangle \left\langle H \right|$&$\frac{1}{{\sqrt 2 }}\left( {\left| {H,+1} \right\rangle  - \left| {H, - 1} \right\rangle } \right)$\\
 $\left|{V,-1}\right\rangle$&$1/2$&$+1$&$\rm{-\pi}/4$&$\rm{\pi}/8$&$\rm{\pi}/4$&$\left| V \right\rangle \left\langle V \right|$&$\frac{1}{{\sqrt 2 }}\left( {\left| {V,+1} \right\rangle  - \left| {V, - 1} \right\rangle } \right)$\\
\end{tabular}
\end{ruledtabular}
\end{table*}

The two-dimensional OAM Hadamard (OH) gate is constructed using a sequence of linear optical elements, including QP, SPP, QWP, half-wave plate (HWP), and polarization beam splitter (PBS). HWP rotated at an angle ${{\pi\mathord{\left/{\vphantom {\pi8}} \right.\kern-\nulldelimiterspace} 8}}$ $[\rm{HWP}({\pi/8})]$ acts on polarization modes as~\cite{Xavier2020ph53, Song2024ph66}
\begin{eqnarray}
&\left| {H,l} \right\rangle  \xrightarrow{\rm{HWP}({\pi/8})} \frac{1}{{\sqrt 2 }}\left( {\left| {H,l} \right\rangle  + \left| {V,l} \right\rangle } \right),\nonumber\\
&\left| {V,l} \right\rangle  \xrightarrow{\rm{HWP}({\pi/8})} \frac{1}{{\sqrt 2 }}\left( {\left| {H,l} \right\rangle  - \left| {V,l} \right\rangle } \right).
\end{eqnarray}
HWP oriented at an angle ${\pi/4 }$ [HWP$({\pi/4})$] induces the swaps $\left| {H,l} \right\rangle   \leftrightarrow \left| {V,l} \right\rangle$. The PBS transmits horizontal polarization state ${\left| {H} \right\rangle }$ and reflects vertical polarization state ${\left| {V} \right\rangle }$. The functions of other linear optical elements have previously been described. The detailed parameter configurations of the linear optical elements used in the two-dimensional OH gate are listed in Table~\ref{tab:1}. The operation of the OH gate on OAM modes can be expressed as
\begin{eqnarray}
&\left| {p,+1} \right\rangle  \xrightarrow{\rm{OH}} \frac{1}{{\sqrt 2 }}\left( {\left| {p,+1} \right\rangle  + \left| {p,-1} \right\rangle } \right),\nonumber\\
&\left| {p,-1} \right\rangle  \xrightarrow{\rm{OH}} \frac{1}{{\sqrt 2 }}\left( {\left| {p,+1} \right\rangle  - \left| {p,-1} \right\rangle } \right).
\end{eqnarray}


\subsection{Single-photon projective measurement}

\begin{figure*}
\includegraphics[scale=0.65]{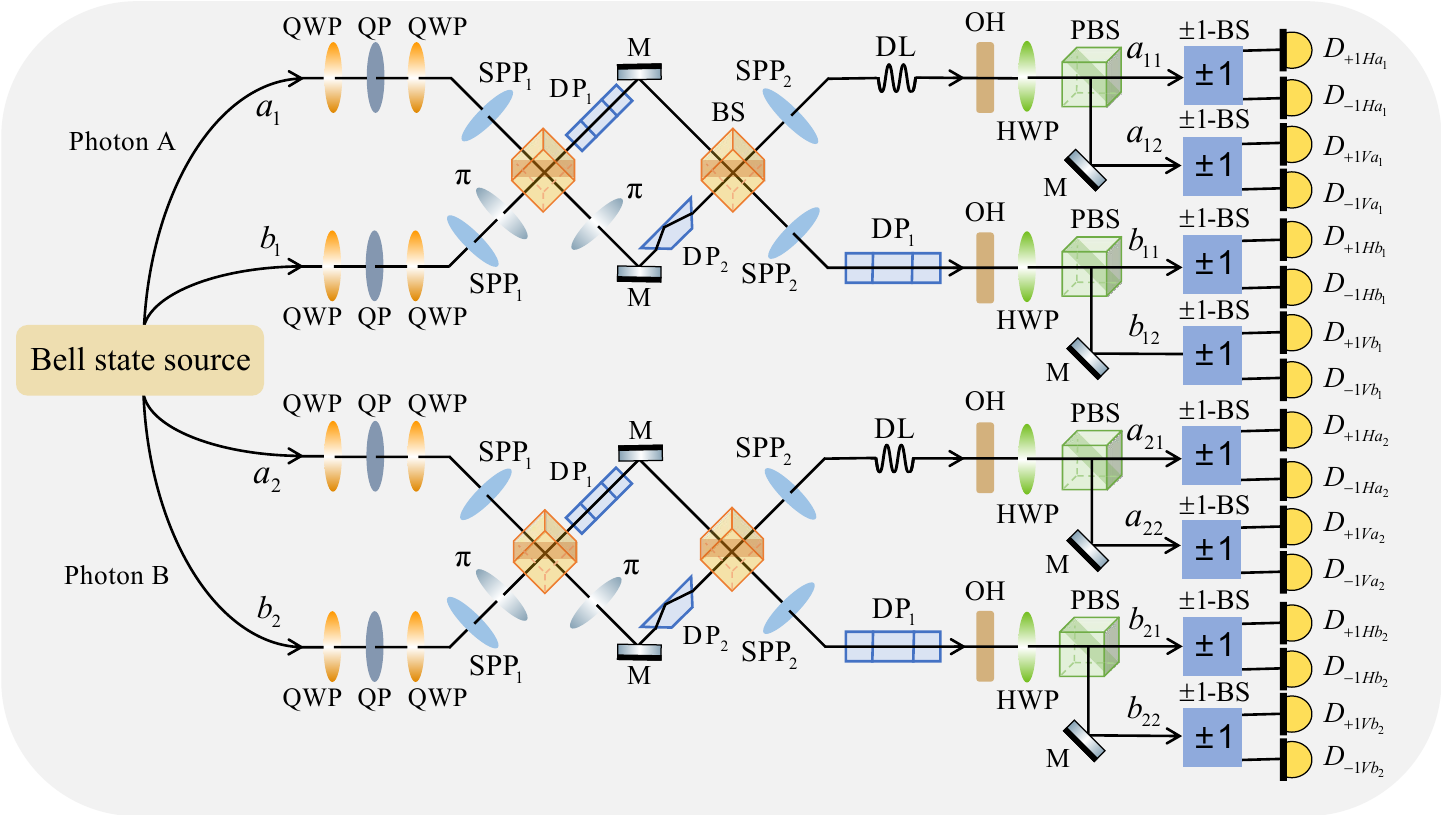}
\caption{\label{fig:2} Schematic diagram of the complete polarized BSA assisted by the auxiliary path and OAM DOFs. QWP denotes a quarter-wave plates oriented at an angle $-\pi/4$. QP denotes a q-plate with topological charges $q=1/2$. SPP$_{1}$ and SPP$_{2}$ denote spiral phase plates with OAM quantum numbers ${+1}$ and ${-1}$, respectively. DP$_{1}$ and DP$_{2}$ denote dove prisms oriented at angles $-\pi/4$ and $0$, respectively. DL is the delay line. HWP denotes a half-wave plate rotated at an angle ${{\pi\mathord{\left/{\vphantom {\pi8}} \right.\kern-\nulldelimiterspace} 8}}$.}
\end{figure*}

The SPPM acting on the polarization, OAM, and path DOFs is used to completely distinguish the output hyperentangled Bell states. The schematic diagram of SPPM is shown in Fig.~\ref{fig:2}(c). In the SPPM implementation, the $\pm$1-BS functions as a single-input, dual-output O-CPS gate, directing incident photons into two distinct paths determined by positive and negative OAM modes. Considering four incident photons passing through a PBS, photons in the states $\left|{H,+1}\right\rangle$ and $\left|{H,-1}\right\rangle$ enter the path $c$ while photons in the states $\left|{V,+1}\right\rangle$ and $\left|{V,-1}\right\rangle$ will enter the path $d$. Subsequent passing through $\pm$1-BS yields definitive detection outcomes: a click in detector $D_{+1H}$ indicates that the incident photon is $\left|{H,+1}\right\rangle$, a click in detector $D_{-1H}$ corresponds to the incident photon $\left|{H,-1}\right\rangle$. Similarly, the input states $\left|{V,+1}\right\rangle$ and $\left|{V,-1}\right\rangle$ trigger clicks in the detectors $D_{+1V}$ and $D_{-1V}$, respectively. This confirms that the quantum information encoded in incident Bell states can be fully extracted using SPPM devices.

\section{Schematic design}\label{sec:3}


The schematic diagram of the complete BSA assisted by auxiliary path and OAM DOFs, is shown in Fig.~\ref{fig:2}. The two-dimensional hyperentangled BSA analyzer comprises P-COS gates controlled by horizontal and vertical polarization modes, O-CPS gates controlled by positive and negative OAM modes, DPs, OH gates, HWPs, and SPPMs acting simultaneously on OAM, polarization and path DOFs. The Bell states to be analyzed, encoded in the polarization DOF, can be written as 
\begin{eqnarray}
{\left| {{\Phi ^ \pm }} \right\rangle _{AB}} = \frac{1}{{\sqrt 2 }}\left( {{{\left| H \right\rangle }_A}{{\left| H \right\rangle }_B} \pm {{\left| V \right\rangle }_A}{{\left| V \right\rangle }_B}} \right),\nonumber\\
{\left| {{\Psi ^ \pm }} \right\rangle _{AB}} = \frac{1}{{\sqrt 2 }}\left( {{{\left| H \right\rangle }_A}{{\left| V \right\rangle }_B} \pm {{\left| V \right\rangle }_A}{{\left| H \right\rangle }_B}} \right), 
\end{eqnarray}
where each is labeled as photon A and photon B.
The auxiliary entanglement is encoded in OAM and path DOFs. The hyperentangled Bell states are injected into the two-dimensional BSA analyzer via paths $a_1a_2$ or paths $b_1b_2$, and can be described as 
\begin{eqnarray}\label{Eq:2}
\left| {\Phi _{in}^ \pm } \right\rangle _{AB} =&& \frac{1}{2}\left( {{{\left| {H,0} \right\rangle }_A}{{\left| {H,0} \right\rangle }_B} \pm {{\left| {V,0} \right\rangle }_A}{{\left| {V,0} \right\rangle }_B}} \right)\nonumber\\
 &&\otimes \left( {a_1a_2 + b_1b_2} \right),\nonumber\\
\left| {\Psi _{in}^ \pm } \right\rangle _{AB} =&& \frac{1}{2}\left( {{{\left| {H,0} \right\rangle }_A}{{\left| {V,0} \right\rangle }_B} \pm {{\left| {V,0} \right\rangle }_A}{{\left| {H,0} \right\rangle }_B}} \right)\nonumber\\
&&\otimes \left( {a_1a_2 + b_1b_2} \right),
\end{eqnarray}
where $a_1$, $a_2$, $b_1$ and $b_2$ represent four different path modes, respectively. After passing through the P-COS gates, the incident Bell states are changed to
\begin{eqnarray}
{\left| {\Phi _1^ \pm } \right\rangle _{AB}}=&&\frac{1}{2}\left( {{{\left| {H, + 1} \right\rangle }_A}{{\left| {H, + 1} \right\rangle }_B} \pm {{\left| {V, - 1} \right\rangle }_A}{{\left| {V, - 1} \right\rangle }_B}} \right)\nonumber\\
&&\otimes \left( {{a_1}{a_2} + {b_1}{b_2}} \right),\nonumber\\
{\left| {\Psi _1^ \pm } \right\rangle _{AB}}=&&
\frac{1}{2}\left( {{{\left| {H, + 1} \right\rangle }_A}{{\left| {V, - 1} \right\rangle }_B} \pm {{\left| {V, - 1} \right\rangle }_A}{{\left| {H, + 1} \right\rangle }_B}} \right) \nonumber\\
&&\otimes \left( {{a_1}{a_2} + {b_1}{b_2}} \right).
\end{eqnarray}
Then the O-CPS gate operations are performed on the states ${\left| {\Phi _1^ \pm } \right\rangle _{AB}}$ and ${\left| {\Psi _1^ \pm } \right\rangle _{AB}}$, these states are converted to
\begin{eqnarray}
{\left| {\Phi _2^ \pm } \right\rangle _{AB}}=&&\frac{1}{2}\left( {{{\left| {H, + 1} \right\rangle }_A}{{\left| {H, + 1} \right\rangle }_B} \pm {{\left| {V, - 1} \right\rangle }_A}{{\left| {V, - 1} \right\rangle }_B}} \right)\nonumber\\
&&\otimes \left( {{a_1}{a_2} + {b_1}{b_2}} \right),\nonumber\\
{\left| {\Psi _2^ \pm } \right\rangle _{AB}}=&&
\frac{1}{2}\left( {{{\left| {H, + 1} \right\rangle }_A}{{\left| {V, - 1} \right\rangle }_B} \pm {{\left| {V, - 1} \right\rangle }_A}{{\left| {H, + 1} \right\rangle }_B}} \right) \nonumber\\
&&\otimes \left( {{a_1}{b_2} + {b_1}{a_2}} \right).
\end{eqnarray}

A DP rotated at an angle $-\pi/4$ [${\rm{DP}(-{\pi/4})}$] operation can be expressed as
\begin{eqnarray}
\left| {H, +1} \right\rangle  \xrightarrow{\rm{DP}({-\pi/4})} \left| {H,-1} \right\rangle ,\nonumber\\
\ \left| {V,-1} \right\rangle  \xrightarrow{\rm{DP({-\pi/4})}} \left| {V, +1} \right\rangle.
\end{eqnarray}
The ${\rm{DPs}({-\pi/4})}$ are performed on the photon $B$ in the paths $b_{12}$ and $b_{22}$, the states ${\left| {\Phi _2^ \pm } \right\rangle _{AB}}$ and ${\left| {\Psi _2^ \pm } \right\rangle _{AB}}$ are transformed into
\begin{eqnarray}
{\left| {\Phi _3^ \pm } \right\rangle _{AB}}=&&\frac{1}{2}\left( {{{\left| {H, + 1} \right\rangle }_A}{{\left| {H, - 1} \right\rangle }_B} \pm {{\left| {V, - 1} \right\rangle }_A}{{\left| {V, + 1} \right\rangle }_B}} \right)\nonumber\\
&&\otimes \left( {{a_1}{a_2} + {b_1}{b_2}} \right),\nonumber\\
{\left| {\Psi _3^ \pm } \right\rangle _{AB}}=&&
\frac{1}{2}\left( {{{\left| {H, + 1} \right\rangle }_A}{{\left| {V, + 1} \right\rangle }_B} \pm {{\left| {V, - 1} \right\rangle }_A}{{\left| {H, -1} \right\rangle }_B}} \right) \nonumber\\
&&\otimes \left( {{a_1}{b_2} + {b_1}{a_2}} \right).
\end{eqnarray}
\begin{table*}[htbp]
\centering
\caption{\label{tab:2}Detector outcomes for two-dimensional BSA scheme assisted by the path and OAM DOFs.}
\begin{ruledtabular}
\begin{tabular}{ccccc}
 Initial Bell states & \multicolumn{4}{c}{Detector outcomes} \\ \hline
\multirow{4}{*}{${\left| {{\Phi ^ + }} \right\rangle _{AB}}$}
&${D_{+1H{a_1}}}{D_{+1H{a_2}}}$&${D_{+1V{a_1}}}{D_{  +1V{a_2}}}$&${D_{+1H{a_1}}}{D_{ - 1V{a_2}}}$&${D_{+1V{a_1}}}{D_{ - 1H{a_2}}}$\\
&${D_{ - 1H{a_1}}}{D_{+1V{a_2}}}$&${D_{ - 1V{a_1}}}{D_{+1H{a_2}}}$&${D_{ - 1H{a_1}}}{D_{ - 1H{a_2}}}$&${D_{ - 1V{a_1}}}{D_{ - 1V{a_2}}}$\\
&${D_{+1H{b_1}}}{D_{+1H{b_2}}}$&${D_{+1V{b_1}}}{D_{+1V{b_2}}}$&${D_{+1H{b_1}}}{D_{ - 1V{b_2}}}$&${D_{+1V{b_1}}}{D_{ - 1H{b_2}}}$\\
&${D_{ - 1H{b_1}}}{D_{+1V{b_2}}}$&${D_{ - 1V{b_1}}}{D_{+1H{b_2}}}$&${D_{ - 1H{b_1}}}{D_{ - 1H{b_2}}}$&${D_{ - 1V{b_1}}}{D_{ - 1V{b_2}}}$\\
\hline

$\multirow{4}{*}{${\left| {{\Phi ^ - }} \right\rangle_{AB}}$}$ 
&${D_{+1H{a_1}}}{D_{+1V{a_2}}}$&${D_{+1V{a_1}}}{D_{+1H{a_2}}}$&${D_{+1H{a_1}}}{D_{ -1H{a_2}}}$&${D_{+1V{a_1}}}{D_{ - 1V{a_2}}}$\\
&${D_{ - 1H{a_1}}}{D_{+1H{a_2}}}$&${D_{ - 1V{a_1}}}{D_{+1V{a_2}}}$&${D_{ - 1H{a_1}}}{D_{ - 1V{a_2}}}$&${D_{ - 1V{a_1}}}{D_{ - 1H{a_2}}}$\\
&${D_{+1H{b_1}}}{D_{+1V{b_2}}}$&${D_{+1V{b_1}}}{D_{+1H{b_2}}}$&${D_{+1H{b_1}}}{D_{ - 1H{b_2}}}$&${D_{+1V{b_1}}}{D_{ - 1V{b_2}}}$\\
&${D_{ - 1H{b_1}}}{D_{+1H{b_2}}}$&${D_{ - 1V{b_1}}}{D_{+1V{b_2}}}$&${D_{ - 1H{b_1}}}{D_{ - 1V{b_2}}}$&${D_{ - 1V{b_1}}}{D_{ - 1H{b_2}}}$\\
\hline

$\multirow{4}{*}{${\left| {{\Psi ^ + }} \right\rangle _{AB}}$}$
&${D_{+1H{a_1}}}{D_{+1H{b_2}}}$&${D_{+1V{a_1}}}{D_{+1V{b_2}}}$&${D_{+1H{a_1}}}{D_{ - 1V{b_2}}}$&${D_{+1V{a_1}}}{D_{ - 1H{b_2}}}$\\
&${D_{ - 1H{a_1}}}{D_{+1V{b_2}}}$&${D_{ - 1V{a_1}}}{D_{+1H{b_2}}}$&${D_{ - 1H{a_1}}}{D_{ - 1H{b_2}}}$&${D_{ - 1V{a_1}}}{D_{ - 1V{b_2}}}$\\
&${D_{+1H{b_1}}}{D_{+1H{a_2}}}$&${D_{+1V{b_1}}}{D_{+1V{a_2}}}$&${D_{+1H{b_1}}}{D_{ - 1V{a_2}}}$&${D_{+1V{b_1}}}{D_{ - 1H{a_2}}}$\\
&${D_{ - 1H{b_1}}}{D_{+1V{a_2}}}$&${D_{ - 1V{b_1}}}{D_{+1H{a_2}}}$&${D_{ - 1H{b_1}}}{D_{ - 1H{a_2}}}$&${D_{ - 1V{b_1}}}{D_{ - 1V{a_2}}}$\\
\hline

$\multirow{4}{*}{${\left| {{\Psi ^ - }} \right\rangle _{AB}}$}$
&${D_{+1H{a_1}}}{D_{+1V{b_2}}}$&${D_{+1V{a_1}}}{D_{+1H{b_2}}}$&${D_{+1H{a_1}}}{D_{ - 1H{b_2}}}$&${D_{+1V{a_1}}}{D_{ - 1V{b_2}}}$\\
&${D_{ - 1H{a_1}}}{D_{+1H{b_2}}}$&${D_{ - 1V{a_1}}}{D_{+1V{b_2}}}$&${D_{ - 1H{a_1}}}{D_{ - 1V{b_2}}}$&${D_{ - 1V{a_1}}}{D_{ - 1H{b_2}}}$\\
&${D_{+1H{b_1}}}{D_{+1V{a_2}}}$&${D_{+1V{b_1}}}{D_{+1H{a_2}}}$&${D_{+1H{b_1}}}{D_{ - 1H{a_2}}}$&${D_{+1V{b_1}}}{D_{ - 1V{a_2}}}$\\
&${D_{ - 1H{b_1}}}{D_{+1H{a_2}}}$&${D_{ - 1V{b_1}}}{D_{+1V{a_2}}}$&${D_{ - 1H{b_1}}}{D_{ - 1V{a_2}}}$&${D_{ - 1V{b_1}}}{D_{ - 1H{a_2}}}$\\
\end{tabular}
\end{ruledtabular}
\end{table*}

The two-dimensional OH gates acting on the OAM modes are performed, which transform the states ${\left| {\Phi _3^ \pm } \right\rangle _{AB}}$ and ${\left| {\Psi _3^ \pm } \right\rangle _{AB}}$ into
\begin{widetext}
\begin{eqnarray}
{\left| {\Phi _4^ + } \right\rangle _{AB}}=&&\frac{1}{4}({\left| {H,+1} \right\rangle _A}{\left| {H,+1} \right\rangle _B} - {\left| {H,+1} \right\rangle _A}{\left| {H, - 1} \right\rangle _B} + {\left| {H, - 1} \right\rangle _A}{\left| {H,+1} \right\rangle _B} - {\left| {H, - 1} \right\rangle _A}{\left| {H, - 1} \right\rangle _B}\nonumber\\
&&+{\left| {V,+1} \right\rangle _A}{\left| {V,+1} \right\rangle _B} + {\left| {V,+1} \right\rangle _A}{\left| {V, - 1} \right\rangle _B} - {\left| {V, - 1} \right\rangle _A}{\left| {V,+1} \right\rangle _B} - {\left| {V, - 1} \right\rangle _A}{\left| {V, - 1} \right\rangle _B})\otimes \left( {{a_1}{a_2} + {b_1}{b_2}} \right),\nonumber\\
{\left| {\Phi _4^ - } \right\rangle _{AB}}=&&\frac{1}{4}({\left| {H,+1} \right\rangle _A}{\left| {H,+1} \right\rangle _B} - {\left| {H,+1} \right\rangle _A}{\left| {H, - 1} \right\rangle _B} + {\left| {H, - 1} \right\rangle _A}{\left| {H,+1} \right\rangle _B} - {\left| {H, - 1} \right\rangle _A}{\left| {H, - 1} \right\rangle _B}\nonumber\\
 &&- {\left| {V,+1} \right\rangle _A}{\left| {V,+1} \right\rangle _B} - {\left| {V,+1} \right\rangle _A}{\left| {V, - 1} \right\rangle _B} + {\left| {V, - 1} \right\rangle _A}{\left| {V,+1} \right\rangle _B} + {\left| {V, - 1} \right\rangle _A}{\left| {V, - 1} \right\rangle _B})\otimes \left( {{a_1}{a_2} + {b_1}{b_2}} \right),\nonumber\\
{\left|{\Psi_{4}^+}\right\rangle_{AB}}=&&\frac{1}{4}({\left| {H,+1} \right\rangle _A}{\left| {V,+1} \right\rangle _B} + {\left| {H,+1} \right\rangle _A}{\left| {V, - 1} \right\rangle _B} + {\left| {H, - 1} \right\rangle _A}{\left| {V,+1} \right\rangle _B} + {\left| {H, - 1} \right\rangle _A}{\left| {V, - 1} \right\rangle _B}\nonumber\\
&&+{\left| {V,+1} \right\rangle _A}{\left| {H,+1} \right\rangle _B} - {\left| {V,+1} \right\rangle _A}{\left| {H, - 1} \right\rangle _B} - {\left| {V, - 1} \right\rangle _A}{\left| {H,+1} \right\rangle _B} + {\left| {V, - 1} \right\rangle _A}{\left| {H, - 1} \right\rangle _B})\otimes({a_1b_2 + b_1a_2}),\nonumber\\
{\left|{\Psi_{4}^-}\right\rangle_{AB}}=&&\frac{1}{4}({\left| {H,+1} \right\rangle _A}{\left| {V,+1} \right\rangle _B} + {\left| {H,+1} \right\rangle _A}{\left| {V, - 1} \right\rangle _B} + {\left| {H, - 1} \right\rangle _A}{\left| {V,+1} \right\rangle _B} + {\left| {H, - 1} \right\rangle _A}{\left| {V, - 1} \right\rangle _B}\nonumber\\
 &&- {\left| {V,+1} \right\rangle _A}{\left| {H,+1} \right\rangle _B} + {\left| {V,+1} \right\rangle _A}{\left| {H, - 1} \right\rangle _B} + {\left| {V, - 1} \right\rangle _A}{\left| {H,+1} \right\rangle _B} - {\left| {V, - 1} \right\rangle _A}{\left| {H, - 1} \right\rangle _B})\otimes({a_1b_2 + b_1a_2}).\nonumber\\
\end{eqnarray}
\end{widetext}
The $\rm{HWP}({\pi/8})$ operations acting on the polarization modes are performed, which transform the states ${\left| {\Phi _4^ + } \right\rangle _{AB}}$, ${\left| {\Phi _4^ - } \right\rangle _{AB}}$, ${\left| {\Psi _4^ - } \right\rangle _{AB}}$ and ${\left| {\Psi _4^ - } \right\rangle _{AB}}$ into
\begin{widetext}
\begin{eqnarray}
{\left| {\Phi _{out}^ + } \right\rangle _{AB}}=&&\frac{1}{4}\left( {{{\left| {H,+1} \right\rangle }_A}{{\left| {H,+1} \right\rangle }_B} + \left| {V,+1} \right\rangle_A \left| {V,+1} \right\rangle_B  - {{\left| {H,+1} \right\rangle }_A}{{\left| {V, - 1} \right\rangle }_B} - {{\left| {V,+1} \right\rangle }_A}{{\left| {H, - 1} \right\rangle }_B}} \right.\nonumber\\
&&\left. { + {{\left| {H, - 1} \right\rangle }_A}{{\left| {V,+1} \right\rangle }_B} + {{\left| {V, - 1} \right\rangle }_A}{{\left| {H,+1} \right\rangle }_B} - {{\left| {H, - 1} \right\rangle }_A}{{\left| {H, - 1} \right\rangle }_B} - {{\left| {V, - 1} \right\rangle }_A}{{\left| {V - 1} \right\rangle }_B}} \right)\otimes\left( {a_1a_2 + b_1b_2} \right),\nonumber\\
{\left| {\Phi _{out}^ - } \right\rangle _{AB}}=&&\frac{1}{4}\left( {{{\left| {H,+1} \right\rangle }_A}{{\left| {V,+1} \right\rangle }_B} + \left| {V,+1} \right\rangle_A \left| {H,+1} \right\rangle_B  - {{\left| {H,+1} \right\rangle }_A}{{\left| {H, - 1} \right\rangle }_B} - {{\left| {V,+1} \right\rangle }_A}{{\left| {V, - 1} \right\rangle }_B}} \right.\nonumber\\
&&\left. { + {{\left| {H, - 1} \right\rangle }_A}{{\left| {H,+1} \right\rangle }_B} + {{\left| {V, - 1} \right\rangle }_A}{{\left| {V,+1} \right\rangle }_B} - {{\left| {H, - 1} \right\rangle }_A}{{\left| {V, - 1} \right\rangle }_B} - {{\left| {V, - 1} \right\rangle }_A}{{\left| {H, - 1} \right\rangle }_B}} \right)\otimes\left( {a_1a_2 + b_1b_2} \right),\nonumber\\
 {\left| {\Psi _{out}^ + } \right\rangle _{AB}}=&&\frac{1}{4}\left( {{{\left| {H,+1} \right\rangle }_A}{{\left| {H,+1} \right\rangle }_B} - {{\left| {V,+1} \right\rangle }_A}{{\left| {V,+1} \right\rangle }_B} - {{\left| {H,+1} \right\rangle }_A}{{\left| {V, - 1} \right\rangle }_B} + {{\left| {V,+1} \right\rangle }_A}{{\left| {H, - 1} \right\rangle }_B}} \right.\nonumber\\
&&\left. { - {{\left| {H, - 1} \right\rangle }_A}{{\left| {V,+1} \right\rangle }_B} + {{\left| {V, - 1} \right\rangle }_A}{{\left| {H,+1} \right\rangle }_B} + {{\left| {H, - 1} \right\rangle }_A}{{\left| {H, - 1} \right\rangle }_B} - {{\left| {V, - 1} \right\rangle }_A}{{\left| {V, - 1} \right\rangle }_B}} \right)\otimes\left( {a_1b_2 + b_1a_2} \right),\nonumber\\
{\left| {\Psi _{out}^ - } \right\rangle _{AB}}=&&\frac{1}{4}\left( {{{\left| {H,+1} \right\rangle }_A}{{\left| {V,+1} \right\rangle }_B} - {{\left| {V,+1} \right\rangle }_A}{{\left| {H,+1} \right\rangle }_B} - {{\left| {H,+1} \right\rangle }_A}{{\left| {H, - 1} \right\rangle }_B} + {{\left| {V,+1} \right\rangle }_A}{{\left| {V, - 1} \right\rangle }_B}} \right.\nonumber\\
&&\left. { - {{\left| {H, - 1} \right\rangle }_A}{{\left| {H,+1} \right\rangle }_B} + {{\left| {V, - 1} \right\rangle }_A}{{\left| {V,+1} \right\rangle }_B} + {{\left| {H, - 1} \right\rangle }_A}{{\left| {V, - 1} \right\rangle }_B} - {{\left| {V, - 1} \right\rangle }_A}{{\left| {H, - 1} \right\rangle }_B}} \right)\otimes\left( {a_1b_2 + b_1a_2} \right).
\end{eqnarray}
\end{widetext}

Finally, complete SPPM acting on the polarization, OAM, and path DOFs are performed to distinguish the output hyper-entanglement Bell states. The different outcomes of projective measurement in polarization, OAM and path modes can be accurately identified. The quantum information of the incident Bell states can be extracted based on the responses of 16 different detectors, whose results are summarized in Table~\ref{tab:2}. Thus, the Bell states encoded in polarization DOF can be completely fully analyzed with the assistance of OAM and path entanglement.

\section{Conclusion}
In summary, we have theoretically proposed a fully deterministic scheme for the complete analysis of polarization-encoded Bell states, using OAM and path DOFs within a linear optical framework. In contrast to previous protocols that rely on inefficient nonlinear processes or external auxiliary quantum resources, our BSA scheme offers several key advantages. First, the linear-optics-based hyper-entangled analyzer eliminates the dependence on nonlinear optical processes, which provides a critical practical bottleneck in experimental implementations. Second, the avoidance of external auxiliary quantum systems simplifies quantum circuit architecture and conserves valuable quantum resources, which enhances operational stability. Third, the intrinsic compatibility between OAM/path modes and polarization-encoding systems enables deterministic BSA with a theoretical success probability of 100\%, which overcomes the 50\% limit of linear optical schemes without auxiliary DOFs. These synergistic advancements position our proposal as a pragmatic and experimentally feasible solution for current photonic quantum technology needs. Furthermore, the proposed architecture exhibits strong scalability for high-dimensional quantum systems and multi-photon interaction scenarios. Looking ahead, our scheme establishes a foundational framework for realizing more robust, efficient, and scalable QIP tasks, and provides a viable pathway toward large-scale photonic quantum network implementations.




\begin{acknowledgments}
This work was supported by the National Natural Science Foundation of China (Grant Nos. 12274053 and 62471089) and by the Fundamental Research Funds for the Central Universities (Grant Nos. 04442024070).
\end{acknowledgments}



\bibliography{apsreference}

\end{document}